\documentstyle[12pt,aasms4]{article}
\begin{document}
%\received{}
%\accepted{}
%\journalid{}{}
%\articleid{}{}
%\paperid{19990215v01}
%\slugcomment{}
\lefthead{M. Takeda et al.}
%\righthead{Arrival Directions of Most Energetic Cosmic Rays}

\title{
Small-scale anisotropy of cosmic rays above 10$^{19}$eV 
observed with the Akeno Giant Air Shower Array
}

\author{
{M. Takeda\altaffilmark{1},}
{N. Hayashida\altaffilmark{1},} 
{K. Honda\altaffilmark{2},}
{N. Inoue\altaffilmark{3},}
{K. Kadota\altaffilmark{4},}
{F. Kakimoto\altaffilmark{4},}		\\
{K. Kamata\altaffilmark{5},}
{S. Kawaguchi\altaffilmark{6},}
{Y. Kawasaki\altaffilmark{7},}
{N. Kawasumi\altaffilmark{8},}
{E. Kusano\altaffilmark{3},}		\\
{Y. Matsubara\altaffilmark{9},}
{K. Murakami\altaffilmark{10},}
{M. Nagano\altaffilmark{11},}
{D. Nishikawa\altaffilmark{1},}
{H. Ohoka\altaffilmark{1},}		\\
{S. Osone\altaffilmark{1},}
{N. Sakaki\altaffilmark{1},}
{M. Sasaki\altaffilmark{1},}
{K. Shinozaki\altaffilmark{3},}
{N. Souma\altaffilmark{3},}
{M. Teshima\altaffilmark{1},}		\\
{R. Torii\altaffilmark{1},}
{I. Tsushima\altaffilmark{8},}
{Y. Uchihori\altaffilmark{12},}
{T. Yamamoto\altaffilmark{1},}
{S. Yoshida\altaffilmark{1},}
{and H. Yoshii\altaffilmark{13}}	\\
}

\altaffiltext{1}{ 
Institute for Cosmic Ray Research, University of Tokyo, Tokyo 188-8502, Japan
}
\altaffiltext{2}{ 
Faculty of Engineering, Yamanashi University, Kofu 400-8511, Japan
}
\altaffiltext{3}{ 
Department of Physics, Saitama University, Urawa 338-8570, Japan
}
\altaffiltext{4}{ 
Department of Physics, Tokyo Institute of Technology, Tokyo 152-8551, Japan
}
\altaffiltext{5}{ 
Nishina Memorial Foundation, Komagome, Tokyo 113-0021, Japan
}
\altaffiltext{6}{ 
Faculty of General Education, Hirosaki University, Hirosaki 036-8560, Japan
}
\altaffiltext{7}{ 
Department of Physics, Osaka City University, Osaka 558-8585, Japan
}
\altaffiltext{8}{ 
Faculty of Education, Yamanashi University, Kofu 400-8510, Japan
}
\altaffiltext{9}{ 
Solar-Terrestrial Environment Laboratory, Nagoya University, 
Nagoya 464-8601, Japan
}
\altaffiltext{10}{ 
Nagoya University of Foreign Studies, Nissin, Aichi 470-0131, Japan
} 
\altaffiltext{11}{ 
Department of Applied Physics and Chemistry, Fukui Institute of Technology, 
Fukui 910-8505, Japan 
}
\altaffiltext{12}{ 
National Institute of Radiological Sciences, Chiba 263-8555, Japan
}
\altaffiltext{13}{ 
Department of Physics, Ehime University, Matsuyama 790-8577, Japan
}

\begin{abstract}
With the Akeno Giant Air Shower Array (AGASA), 
581 cosmic rays above 10$^{19}$eV, 47 above 4 $\times$ 10$^{19}$eV, 
and 7 above 10$^{20}$eV are observed until August 1998. 
Arrival direction distribution of these extremely high energy 
cosmic rays has been studied. 
While no significant large-scale anisotropy is found 
on the celestial sphere, 
some interesting clusters of cosmic rays are observed. 
Above 4 $\times$ 10$^{19}$eV, 
there are one triplet and three doublets 
within separation angle of 2.5$\arcdeg$ and 
the probability of observing these clusters by a chance coincidence 
under an isotropic distribution is smaller than 1 \%. 
Especially the triplet is observed against expected 0.05 events. 
The $\cos(\theta_{GC})$ distribution expected 
from the Dark Matter Halo model fits the data as well as 
an isotropic distribution 
above 2 $\times$ 10$^{19}$eV and 4 $\times$ 10$^{19}$eV, 
but is a poorer fit than isotropy above 10$^{19}$eV. 
Arrival direction distribution of seven 10$^{20}$eV cosmic rays 
is consistent with that of lower energy cosmic rays 
and is uniform. 
Three of seven are members of doublets above about 4 $\times$ 10$^{19}$eV. 
\end{abstract}


%*********************************************************************
\section{Introduction}
\label{sect:intro}

Investigation on anisotropy of extremely high energy cosmic rays is 
one of the most important aspects to reveal their origin. 
In energies $\gtrsim$ 10$^{19}$eV, 
cosmic rays slightly deflect in the galactic magnetic field 
if they are protons of galactic origin, 
so that one could observe the correlation of their arrival directions 
with the galactic structure. 
Especially in the highest observed energy range, 
correlation of cosmic rays with the local structure of galaxies 
may be expected if their origins are nearby astrophysical objects 
and the intergalactic magnetic field is less than 10$^{-9}$ gauss. 

In the 1980's, 
Wdowczyk, Wolfendale and their collaborators (\cite{WW84,szabelski86a}) 
have shown that 
excess of cosmic rays from the direction of the galactic plane 
increases systematically with energy until a little above 10$^{19}$eV, 
though the available data was not statistically enough at that time. 
Gillman and Watson (1993) have summarized 
anisotropies in right ascension and galactic latitude 
combining the Haverah Park data set with 
the data sets from the arrays at 
Volcano Ranch (\cite{linsley80a}),
Sydney (\cite{winn86a}) and Yakutsk (\cite{efimov88a}). 
No convincing anisotropies were observed; but 
large amplitude of the second harmonics 
at (4 -- 8) $\times$ 10$^{18}$eV was reported. 
Ivanov (1998) showed, with the Yakutsk data set,  
a north-south asymmetry in the galactic latitude distribution 
which is the southern excess with 3.5 $\sigma$ deviation 
from an isotropic distribution 
in (5 -- 20) $\times$ 10$^{18}$eV. 

Recently, we have shown a significant anisotropy 
with first harmonic amplitude of $\simeq$ 4 \% 
in (0.8 -- 2.0) $\times$ 10$^{18}$eV, 
which corresponds to the chance probability of 0.2 \% 
due to fluctuation of an isotropic distribution 
(\cite{kusano98a}). 
This anisotropy shows broad cosmic-ray flow 
from the directions of the galactic center and the Cygnus regions. 
In the higher energies, 
no significant large-scale anisotropy was found. 
Bird et al. (1998) have shown 
the galactic plane enhancement in the similar energy range. 
These experiments show that 
significant fraction of cosmic rays around 10$^{18}$eV 
come from galactic sources. 

In the much higher energy range $\geq$ 4 $\times$ 10$^{19}$eV, 
Stanev et al. (1995) have claimed that 
cosmic rays exhibit a correlation with the direction of 
the supergalactic plane 
and the magnitude of the observed excess is 2.5 -- 2.8 $\sigma$ 
in terms of Gaussian probabilities. 
Their result was mainly based on the Haverah Park data set. 
In the same energy range, 
such large-scale correlation with the supergalactic plane 
was not observed in the data sets of the 
AGASA (\cite{uchihori96a}), 
SUGAR (\cite{kewley96a}) 
and Fly's Eye (\cite{bird98a}) experiments.
However, AGASA observed three pairs of cosmic rays 
above 4 $\times$ 10$^{19}$eV 
within a limited solid angle of the experimental accuracy 
and the chance probability is 2.9 \% 
if cosmic rays distribute uniformly in the AGASA field of view. 
Two out of three are located nearly on the supergalactic plane. 
If cosmic rays in each of these pairs come from the same source, 
the detailed study on energy, arrival time and direction distribution 
of these clusters may bring information on their source and 
the intergalactic magnetic field (\cite{sigl98a,tanco98a}). 

In the observed energy spectrum, 
there are two distinctive energies: $E$ $\simeq$ 10$^{19}$eV 
and 4 $\times$ 10$^{19}$eV. 
The former is the energy where the spectral slope changes 
(\cite{lawrence91a,efimov91a,bird94a,yoshida95a,takeda98a}). 
This is interpreted as 
transition from galactic to extragalactic origin. 
The latter is the energy where the GZK effect (\cite{g66a,zk66a}), 
which is a series of energy loss through interaction 
with the cosmic microwave background photons, 
becomes important on their propagation from sources. 
It is important to study whether the arrival direction distribution 
of cosmic rays changes at these energies. 

Recent result of the AGASA energy spectrum shows 
the extension beyond the expected GZK cutoff (\cite{takeda98a}). 
Since the distance to sources of cosmic rays above 
the expected GZK cutoff is limited to 50 Mpc 
(\cite{hill85a,berezinsky88a,yoshida93a}), 
their arrival directions may be correlated 
with luminous matter distribution  
if they are astrophysical source origin such as 
hot spots of radio galaxies 
(\cite{biermann87a,takahara90a,rachen93a,ostrowski98a}), 
active galactic nuclei 
(\cite{blandford76a,lovelace76a,rees82a}), 
accretion flow to a cluster of galaxies 
(\cite{kang97a}), 
relativistic shocks in gamma-ray bursts
(\cite{vietri95a,waxmann95a}), 
and so on. 
There is another possibility that 
most energetic cosmic rays are generated through 
decay of supermassive ``X'' particles 
related to topological defects (\cite{bhatt98a}, reference therein). 
In this case, 
arrival directions of most energetic cosmic rays 
are not necessarily associated with luminous matters. 
If such particles are the part of Dark Matter and 
are concentrated in the galactic halo, 
anisotropy associated with our galactic halo is expected 
(\cite{kuzmin97a,berezinsky97a}). 

In this paper, 
we first examine large-scale anisotropy 
in terms of various coordinates 
using the data set of the Akeno Giant Air Shower Array (AGASA) 
until August 1998, 
including the old data set of the Akeno 20 km$^{2}$ array (A20) before 1990. 
Then we search for the small-scale anisotropy above 10$^{19}$eV 
with the AGASA data set. 

%*****************************************************************************
\section{Experiment}
\label{sect:exp}

The Akeno Observatory is situated 
at 138$\arcdeg$ 30$\arcmin$ E and 35$\arcdeg$ 47$\arcmin$ N. 
AGASA consists of 111 surface detectors deployed over 
an area of about 100 km$^{2}$, 
and has been in operation since 1990 (\cite{chiba92a,ohoka97a}). 
A20 is a prototype detector system of AGASA, 
operated from 1984 to 1990 (\cite{teshima86a}), 
and is a part of AGASA after 1990. 

Each surface detector consists of plastic scintillators 
of 2.2 m$^{2}$ area. 
The detectors are placed with a separation of about 1 km. 
They are controlled and operated from a central computer 
through optical fiber network. 
Relative time difference among the detectors are measured 
with 40 nsec accuracy; 
all clocks at detector sites are synchronized 
to the central clock and 
signal-propagation time in cables and electronic devices are 
regularly measured at start of each run (twice a day). 
The details of the AGASA instruments have been 
described in Chiba et al. (1992) and Ohoka et al. (1997).

The accuracy on determination of shower parameters are 
evaluated through the analysis of a large number of artificial events. 
These artificial events are generated with taking account of 
air shower features and fluctuation determined experimentally. 
Figure \ref{fig:acc_dir} shows the accuracy 
on arrival direction determination for cosmic-ray induced air showers 
as a function of energies. 
The vertical axis denotes the opening angle $\Delta \theta$ 
between input (simulated) and output (analyzed) arrival directions. 
The opening angles including 68 \% and 90 \% of data are plotted. 
By analyzing artificial events with the same algorithm used above, 
the accuracy on energy determination is 
estimated to be $\pm$ 30 \% above 10$^{19}$eV (\cite{yoshida95a}).

Table \ref{tbl:num_events} lists the number of selected events, N(E), 
with zenith angles smaller than 45$\arcdeg$ and 
with core locations inside the array area. 
Events below 10$^{19}$eV are used only a reference analysis in this paper. 
The difference 
of N(E $\geq$ 3.2 $\times$ 10$^{19}$eV) $/$ N(E $\geq$ 10$^{19}$eV) 
between A20 and AGASA 
arises from the difference of detection efficiency of each system. 
Seven events are observed above 10$^{20}$eV, 
including one event after Takeda et al. (1998).

%*****************************************************************************
\section{Results}
\label{sect:analysis}

Figure \ref{fig:equ1900}(a) shows arrival directions of cosmic rays 
with energies above 10$^{19}$eV 
on the equatorial coordinates. 
Dots, open circles, and open squares represent cosmic rays 
with energies of (1 -- 4) $\times$ 10$^{19}$eV, 
(4 -- 10) $\times$ 10$^{19}$eV, and $\geq$ 10$^{20}$eV, respectively. 
The shaded regions indicate the celestial regions 
excluded in this paper 
due to the zenith angle cut of $\leq$ 45$\arcdeg$. 
The galactic and supergalactic planes are drawn by the dashed lines. 
``GC'' designates the galactic center. 
Figure \ref{fig:equ1900}(b) shows arrival directions of cosmic rays 
only above 4 $\times$ 10$^{19}$eV on the galactic coordinates. 
Details of the cosmic rays above 4 $\times$ 10$^{19}$eV are listed in 
Table \ref{tbl:40EeV}.

%*****************************************************************************
\subsection{Analysis in the Equatorial Coordinates} 
\label{ssect:equ}

\subsubsection{Harmonic Analysis}
\label{ssect:harmo}

In order to search for cosmic ray anisotropy, 
it is required to compare observed and expected event frequencies 
at each region. 
An expected frequency is easily estimated as far as the exposure 
in each direction can be obtained; 
the uniformity of observation time on solar time for several years, 
which results in the uniform observation in right ascension, 
is expected for a surface array detection system 
operating in stable like AGASA. 
The fluctuation of the observation time on the local sidereal time 
is (0.2 $\pm$ 0.1) \% 
which is small enough compared with anisotropy in this energy range, 
so that the exposure (observation time $\times$ collection area) 
in right ascension is quite uniform.

Figure \ref{fig:harmo} shows results of the first (left) 
and second (right) harmonics in right ascension. 
The amplitude (top), the phase (middle), 
and the chance probability (bottom) are shown in each energy bin. 
In the top panels of the harmonic amplitude, 
the shaded region is expected from statistical fluctuation of 
an isotropic distribution with the chance probability larger than 10 \%. 
No significant anisotropy above this level is found 
above 3.2 $\times$ 10$^{18}$eV. 
This is consistent with our previous paper (\cite{kusano98a}), 
in which zenith angles up to 60$\arcdeg$ were used.

\subsubsection{Declination Distribution}
\label{ssect:dec}

Figure \ref{fig:dec1900} shows 
the declination distribution of events 
above 10$^{19}$eV (light shaded histogram) and 
10$^{20}$eV (dark shaded histogram). 
A solid curve is a third order polynomial function 
fitted to the light shaded histogram. 
This curve is consistent with 
the zenith angle dependence of the AGASA exposure and 
considered to be the expected distribution 
if cosmic rays distribute isotropically on the celestial sphere. 
Since the trigger efficiency is independent of 
energy above 10$^{19}$eV and zenith angle less than 45$\arcdeg$, 
this distribution is applied to in higher energies. 
Excess with 2.5 $\sigma$ deviation is found 
in $\delta$ $=$ [30$\arcdeg$, 40$\arcdeg$] 
and this will be discussed later. 

%=====	119 / {(119 - 94.36) / sqrt(94.36)) = + 2.536569 \sigma
%=====		Li&Ma: +1.698764 \sigma

%*****************************************************************************
\subsection{Analysis in the Galactic and Supergalactic Coordinates} 
\label{ssect:gal_sgl}

\subsubsection{Galactic and Supergalactic Plane Enhancement} 
\label{ssect:fe}

If cosmic rays have origin associating with nearby astrophysical objects, 
we may expect cosmic-ray anisotropy correlated with the galactic or 
supergalactic plane. 
Figure \ref{fig:lat_gsg} shows the latitude distribution on 
the galactic (left) and supergalactic (right) coordinates 
in three energy ranges of 
(1 -- 2) $\times$ 10$^{19}$eV (top), 
(2 -- 4) $\times$ 10$^{19}$eV (middle), 
and $\geq$ 4 $\times$ 10$^{19}$eV (bottom). 
A solid line in each panel indicates the cosmic-ray intensity expected 
from an isotropic distribution. 
In order to examine any preference for arrival directions along 
the galactic and supergalactic planes, 
the plane enhancement parameter $ f_{E} $ 
introduced by Wdowczyk and Wolfendale (1984) 
was used. 
The $ f_{E} $ value characterizes the anisotropy expressed by: 
\begin{equation}
	I_{obs}(b) / I_{exp}(b) = (1 - f_{E}) + 
		1.402 \, f_{E} \, \exp(-b^2), 
\end{equation}
where $ b $ is galactic or supergalactic latitude in radians, 
$ I_{obs}(b) $ and $ I_{exp}(b) $ are 
observed and expected intensities at latitude $ b $. 
A positive $ f_{E} $ value suggests a galactic or supergalactic 
plane enhancement, 
$ f_{E} = 0 $ indicates that arrival direction distribution is isotropic, 
and a negative $ f_{E} $ shows depression around the plane. 
Figure \ref{fig:fe_gsg} shows the dependence of $ f_{E} $ 
on the primary energy 
for the galactic (left) and supergalactic (right) coordinates. 
Some excess can be seen around the supergalactic plane 
in the seventh energy bin ($\log$(E[eV]) $=$ [19.1, 19.2]), 
where $ f_{E}^{SG} = 0.36 \pm 0.15 $. 
%=====	2.4728836 \sigma
In other energies, 
the arrival direction distribution is consistent with 
an isotropic distribution.

\subsubsection{$\theta_{GC}$ Distribution}
\label{ssect:gc_sun}

Figure \ref{fig:axhist_GC} shows the $\cos$($\theta_{GC}$) distribution, 
where $\theta_{GC}$ is the opening angle between 
the cosmic-ray arrival direction and the galactic center direction, 
with energies above 10$^{19}$eV (top), 
2 $\times$ 10$^{19}$eV (middle), and 4 $\times$ 10$^{19}$eV (bottom). 
Histograms are the observed distribution and 
the solid curves are expected from an isotropic distribution. 
The observed distribution is consistent with the solid curve 
in all energy ranges. 
The dashed and dotted curves are expected from 
the Dark Matter Halo model (\cite{berezinsky98b}) and 
will be discussed in Section \ref{ssect:berezinsky}.

%*****************************************************************************
\subsection{Significance Map of Cosmic-Ray Excess/Deficit}
\label{ssect:sky_map}

There is no statistically significant large-scale anisotropy 
in the above one-dimensional analyses. 
Here, we search for two-dimensional anisotropy 
with taking account of the angular resolution event by event.

Figures \ref{fig:dmap1900} and \ref{fig:dmap1960} show 
the contour maps of the cosmic-ray excess or deficit 
with respect to an isotropic distribution 
above 10$^{19}$eV and 4 $\times$ 10$^{19}$eV, respectively. 
A bright region indicates that 
the observed cosmic-ray intensity is larger than 
the expected intensity 
and a dark region shows a deficit region. 
For each observed event, we calculate a point spread function 
which is assumed to be a normalized Gaussian probability distribution 
with a standard deviation of the angular resolution $ \Delta \theta $ 
obtained from Figure \ref{fig:acc_dir}. 
The probability densities of all events are folded into cells 
of 1$\arcdeg$ $\times$ 1$\arcdeg$ in the equatorial coordinates. 
At each cell, we sum up densities within 
4.0$\arcdeg$ radius for Figure \ref{fig:dmap1900} and
2.5$\arcdeg$        for Figure \ref{fig:dmap1960}. 
These radii are obtained from $ \sqrt{2} \times \Delta \theta $, 
and they would make excess regions clearer. 
The reference distribution is obtained 
from an isotropic distribution. 
In these figures, small statistics of observed and expected events 
result in bright regions at the lower and higher declination and 
hence bright spots below $\delta = 0\arcdeg$ are not significant. 
Two distinctive bright regions are found in Figure \ref{fig:dmap1900}, 
which are broader than the angular resolution. 
They are referred to as broad clusters, such as 
the BC1 (20$^{h}$50$^{m}$, 32$\arcdeg$) and 
    BC2 ( 1$^{h}$40$^{m}$, 35$\arcdeg$). 
The member events within 4$\arcdeg$ radius of BC1 are listed in 
Table \ref{tbl:clst_1900}. 
Four brighter regions in the middle declination are found 
in Figure \ref{fig:dmap1960}: 
the C1 -- C4 clusters 
which are noted in the eighth column of Table \ref{tbl:40EeV}. 
The C1 -- C3 clusters follow 
the notation used in our previous analysis (\cite{uchihori96a}). 
The C2 cluster is observed in both energy ranges. 

In Figure \ref{fig:dmap1900}, the contour map 
has eight steps in [$-3\sigma$, $+3\sigma$]; 
lower two steps below $-1.5\sigma$ are absent. 
The significance of deviation from an isotropic distribution 
are estimated to be 
2.4 $\sigma$ at the  C2 cluster, 
2.7 $\sigma$ at the BC1 cluster, and
2.8 $\sigma$ at the BC2 cluster. 
The arrival directions of cosmic rays around the BC1 cluster 
are shown in Figure \ref{fig:clst_c0}(a), 
and a radius of each circle corresponds to the logarithm of its energy. 
Shaded circles have energies above 10$^{19}$eV and 
open circles below 10$^{19}$eV. 
Figure \ref{fig:clst_c0}(b) shows the arrival time -- energy relation, 
and open circles denote members of the BC1 cluster. 
The members of the BC1 cluster have energies 
between 10$^{19}$eV and 2.5 $\times$ 10$^{19}$eV 
and no excess of cosmic rays are observed 
below 10$^{19}$eV around this direction. 
Five members of the BC1 cluster are observed 
around MJD 50,000. 
This cluster is in the direction of 
a famous supernova remnant --- the Cygnus Loop 
which extends about 3$\arcdeg$ 
around (20$^{h}$50$^{m}$, 30$\arcdeg$ 34$\arcmin$). 
The BC2 cluster is the broader cluster without a clear boundary. 
The BC1 and BC2 clusters contribute the excess 
around $\delta = 35\arcdeg$ shown in Figure \ref{fig:dec1900}. 
The C2 and BC2 clusters are located near the supergalactic plane 
and lead the largest $ f_{E}^{SG} $ value in Section \ref{ssect:fe}. 

For small statistics of observed events, 
Figure \ref{fig:dmap1960} reflects 
the arrival directions of individual events 
(open squares and open circles in Figure \ref{fig:equ1900}). 
The brightest peak is at the C2 cluster where 
three cosmic rays are observed against expected 0.05 events. 
It is possible that some of these clusters are observed 
by a chance coincidence. 
It should be noted, however, that two of these clusters 
--- the doublet (C1) including the AGASA highest energy event and 
the triplet (C2) --- 
lie near the supergalactic plane, 
as pointed in our previous analysis (\cite{uchihori96a}). 
The arrival directions (left) and arrival time -- energy relation (right) 
for the C1 (top) and C2 (bottom) clusters are shown 
in Figure \ref{fig:clst_1960}. 
A radius of each circle in the left panels 
corresponds to the logarithm of its energy, and 
open circles in the right panels denote members of the C1 and C2 clusters. 
Around the C2 cluster, several lower energy cosmic rays are observed 
very close to the C2 cluster.

%*****************************************************************************
\subsection{Cluster Analysis}
\label{ssect:cluster}

The threshold energy of 4 $\times$ 10$^{19}$eV is 
one distinctive energy where the GZK effect becomes large 
as mentioned in Section \ref{sect:intro}. 
It is, however, quite important to examine 
what kind of dependence on threshold energy is operating. 

To begin with, we estimate the chance probability of observing 
one triplet and three doublets 
from 47 cosmic rays above 4 $\times$ 10$^{19}$eV. 
A cluster of cosmic rays is defined as follows:
\begin{enumerate}
\item	Define the $i$-th event;
\item	Count the number of events within a circle of
	radius 2.5$\arcdeg$ centered on the arrival direction
        of the $i$-th event;
\item	If this number of events exceeds a certain 
	threshold value $ N_{th} $, 
	the $i$-th event is counted as a cluster.
\end{enumerate}
This procedure was repeated for total 47 events 
and then the total number of clusters with $N_{th}$ was determined. 
The chance probability $P_{ch}$ of observing
this number of clusters under an isotropic distribution 
is obtained from the distribution of the number of clusters 
using 10,000 simulated data sets.
These simulated data sets were also analyzed 
by the same procedure described above. 
Out of 10,000 simulations, 32 trials had equal or more 
doublets ($N_{th} = 2$) than the observed data set, 
so that $ P_{ch} = 0.32 \% $. 
And $ P_{ch} = 0.87 \% $ for triplets ($N_{th} = 3$).

Then, the energy dependence for observing (a) doublets and (b) triplets 
are estimated and the results are shown in Figure \ref{fig:clst_edep}. 
When a new cluster is added above a threshold energy, 
a histogram changes discontinuously at that energy. 
At the maximum threshold energy where the triplet is detected, 
we find $P_{ch}$ $=$ 0.16 \% in Figure \ref{fig:clst_edep}(b). 
The narrow peaks of $P_{ch}$ $\simeq$ 0.1 \% 
above 4 $\times$ 10$^{19}$eV in Figures \ref{fig:clst_edep}(a) 
result from the C1, C3 and C4 doublets, 
and another doublet C5 is found just below 4 $\times$ 10$^{19}$eV. 
Here, these chance probabilities are estimated without 
taking the degree of freedom on the threshold energy into account. 
However, the chance probabilities 
are smaller than 1 \% and 
don't vary abruptly with energies above 4 $\times$ 10$^{19}$eV. 
This means that the threshold energy of 4 $\times$ 10$^{19}$eV 
for doublet and triplet in Figure \ref{fig:dmap1960} 
may indicate any critical energy, 
and suggests that their sources are not very far 
being different from those below this energy. 
%The triplet coincidence is therefore independent of the threshold energy. 

%*****************************************************************************
\subsection{10$^{20}$eV Events}
\label{ssect:100EeV}

Seven events have been observed with energies above 10$^{20}$eV, 
and their energies and coordinates are also listed 
in Table \ref{tbl:40EeV}. 
Their declination are near $\delta \simeq 20\arcdeg$ 
while an isotropic distribution is shown by the solid curve 
in Figure \ref{fig:dec1900}. 
To check whether these seven events distribute isotropically or not, 
we compare celestial distribution of seven 10$^{20}$eV events 
with that for events between 10$^{19}$eV and 10$^{20}$eV 
in ten different coordinates. 
The Kolmogorov-Smirnov (KS) test (\cite{numerical}) was used 
for avoiding any binning effect. 
The results are summarized in Table \ref{tbl:100EeV_KS}. 
The smallest KS probability in Table \ref{tbl:100EeV_KS} is 
2.5 \% for the declination distribution; but 
this probability becomes larger using data set 
above 6.3 $\times$ 10$^{19}$eV. 
One interesting feature is that five 10$^{20}$eV cosmic rays 
come from south-west of the AGASA array, 
where the strength of the geomagnetic field component 
which is perpendicular to an air shower axis is larger 
than the other directions (\cite{stanev97a}).

%*****************************************************************************
\section{Discussion}
\label{sect:discuss}

%*****************************************************************************
\subsection{Comparison with other experiments}
\label{ssect:comparison}

Above 3 $\times$ 10$^{18}$eV, 
no large-scale anisotropy has been found with 
the harmonic analysis and the $f_{E}^{G}$ fit. 
Gillman and Watson (1993) summarized the $f_{E}^{G}$ values 
using the data sets obtained mainly from the Haverah Park experiment. 
They obtained no significant deviation from $f_{E}^{G} = 0$. 
The result from the Fly's Eye experiment (\cite{bird98a}) 
is consistent with an isotropic distribution of cosmic rays 
with E $\gtrsim$ 10$^{19}$eV. 
The analysis with the Yakutsk data set (\cite{ivanov97a}) shows 
no significant galactic plane enhancement above 10$^{18}$eV. 
The results from all experiments are consistent with 
this work on no correlation of cosmic rays above 10$^{19}$eV 
with the galactic plane. 
This may implicate the extragalactic origin of cosmic rays 
above 10$^{19}$eV if they are mostly protons. 

The BC1, BC2 and C1 -- C5 clusters are found 
with energies $\geq$ 10$^{19}$eV or $\gtrsim$ 4 $\times$ 10$^{19}$eV. 
The C2 and BC2 clusters lead 
the small preference along the supergalactic plane 
in the energy range of $\log$(E[eV]) $=$ [19.1, 19.2]. 
With the data sets of Haverah Park, Yakutsk, 
and Valcano Ranch (\cite{uchihori96b}) and AGASA, 
another triplet is found at the position of the C1 cluster 
within experimental error box on arrival direction determination. 
This triplet at the C1 cluster position includes 
the AGASA highest energy event and a 10$^{20}$eV Haverah Park event. 
It should be noted that these triplets at the C1 and C2 positions 
are close to the supergalactic plane.

%*****************************************************************************
\subsection{Correlation with Galactic Halo}
\label{ssect:berezinsky}

Kuzmin and Rubakov (1997) and Berezinsky et al. (1997) have suggested 
a cosmic-ray source model associated with Dark Matter distribution 
in our galactic halo. 
In this model, 
most energetic cosmic rays are generated through 
decay of supermassive particles 
which are trapped in the galactic halo and 
thus distribute symmetrically around the galactic center. 
The arrival directions of most energetic cosmic rays, therefore, 
exhibit anisotropy at the Earth (\cite{berezinsky98a}). 
From recent studies by Berezinsky and Mikhailov (1998) and 
Medina Tanco and Watson (1998), 
a significant anisotropy would be expected 
in the first harmonics of right ascension distribution, 
the amplitude of 40 \% at phase about 250$\arcdeg$, 
which is independent of the ISO and NFW models 
of dark matter distribution in the galactic halo. 
The ISO and NFW models are 
described in Kravtsov et al. (1997) and 
Navarro, Frenk and White (1996), respectively. 
This expected anisotropy is consistent with the results of 
the harmonic analysis above 4 $\times$ 10$^{19}$eV 
as shown in Figure \ref{fig:harmo}. 
However, this amplitude is explained with statistical fluctuation 
of an isotropic distribution. 

As shown by the dashed and dotted curves in Figure \ref{fig:axhist_GC}, 
the ISO and NFW models of Dark Matter distribution in the galactic halo 
lead excess toward the galactic center. 
Table \ref{tbl:gal_halo} shows the reduced-$\chi^{2}$ values of 
the observed $\cos(\theta_{GC})$ distribution with the isotropic, 
ISO and NFW models. 
Although the distribution expected from the ISO and NFW models are quite 
different from the observed distribution in energies above 10$^{19}$eV, 
the reduced-$\chi^{2}$ values are close to one another 
above 2 $\times$ 10$^{19}$eV and 4 $\times$ 10$^{19}$eV. 
Above 2 $\times$ 10$^{19}$eV, all three models are acceptable and 
it is hard to distinguish one from another.

%*****************************************************************************
\subsection{Correlation with Nearby Galaxies}
\label{ssect:tanco}

In Section \ref{ssect:cluster}, 
we calculated the chance probability of observing clusters 
under an isotropic distribution. 
If cosmic rays are astrophysical source origin, 
the non-uniform distribution of galaxies or luminous matters 
should be taken into account, 
as claimed by Medina Tanco (1998). 
He calculated trajectories of cosmic rays above 4 $\times$ 10$^{19}$eV 
in the intergalactic magnetic field under the assumption that 
flux of cosmic rays is proportional to the local density of galaxies. 
The expected distribution of cosmic-ray intensity is no more uniform 
and this may result in a strong anisotropy. 
This is different from the results in this paper 
so that our estimation of the chance probability of observing clusters 
under an isotropic distribution is experimentally reliable. 
However, his calculation shows important results: 
the C2 cluster is on top of a maximum of the arrival probability 
for sources located between 20 and 50 Mpc; and 
the C1 cluster locates on a high arrival probability region 
for sources at more than 50 Mpc. 
This suggests the possibility that 
the members of these clusters are generated at different sources. 
One need accumulate further statistics 
to make arrival direction, time and energy relation to be clear 
(\cite{tanco98a,sigl98a})
to distinguish whether the members of clusters come 
from a single source or unrelated sources.

%*****************************************************************************
\subsection{Correlation with the Known Astrophysical Objects}
\label{ssect:correlate}

As mentioned in Section \ref{ssect:cluster}, 
the BC1 cluster is in the direction of the Cygnus Loop (NGC6992/95). 
From the Hillas confinement condition of (magnetic field $\times$ 
size) for cosmic ray acceleration (\cite{hillas84a}), 
the magnetic field in the shock of the Cygnus Loop is too small 
to accelerate cosmic rays up to 10$^{19}$eV. 
And the observed energy distribution and 
bunch of arrival time of the cluster members 
don't favor the diffusive shock acceleration. 
Another possible candidate is PSR 2053$+$36 
with the period of 0.2215 sec 
and the magnetic field of about 3 $\times$ 10$^{11}$ gauss 
(\cite{manchester81a}). 
It may be plausible that 
such highly magnetized pulsar has accelerated cosmic rays 
up to 10$^{19}$eV within a short time (\cite{gunn,goldreigh}). 
It is highly desired to search for any signals 
from this direction in other energy range around MJD 50,000. 

For the C1 -- C5 clusters and 10$^{20}$eV cosmic rays, 
coincidence with known astrophysical objects are searched for 
from three catalogs which are 
the second EGRET catalog (\cite{thompson95a,thompson96a}), 
the CfA redshift catalog (\cite{huchra95a}), and 
the eighth extragalactic redshift catalog (\cite{veron-cetty98a}). 
The selection criteria are the following: 
(i) the separation angles within 4.0$\arcdeg$ 
from a member of each cluster, 
and 2.5$\arcdeg$ for the 10$^{20}$eV cosmic ray; 
(ii) the redshift within 0.02. 
In the CfA catalog, only QSOs/AGNs are selected. 
Candidate objects are listed in Table \ref{tbl:correlate}. 
Out of these objects, Mrk 40 (VV 141, Arp 151) is an interacting galaxy 
and may be most interesting. 
It should be noted that Al-Dargazelli et al. (1996) claimed
that nearby colliding galaxies are most favored as the
sources of clusters (regions of excess events) defined by
them using the world data available before 1996.

%*****************************************************************************
\section{Summary}
\label{sect:summary}

In conclusion, there is no statistically significant 
large-scale anisotropy related to the galactic nor supergalactic plane. 
The slight supergalactic plane enhancement is observed 
just above 10$^{19}$eV 
and arises mainly from the BC2 and C2 clusters. 
Above 4 $\times$ 10$^{19}$eV, 
one triplet and three doublets are found 
and the probability of observing these clusters 
by a chance coincidence is smaller than 1 \%. 
Especially the triplet is observed against expected 0.05 events. 
Out of these clusters, the C2 (AGASA triplet) and 
C1 (doublet including the AGASA highest energy event 
or triplet together with the Haverah Park 10$^{20}$eV event) 
clusters are most interesting; 
they are triplets found in the world data sets 
and are located near the supergalactic plane. 
One should wait for the further high-rate observation 
to distinguish whether the members of clusters come 
from a single source or different sources. 
The $\cos(\theta_{GC})$ distribution expected 
from the Dark Matter Halo model fits the data as well as 
an isotropic distribution 
above 2 $\times$ 10$^{19}$eV and 4 $\times$ 10$^{19}$eV, 
but is a poorer fit than isotropy above 10$^{19}$eV. 
The arrival direction distribution of the 10$^{20}$eV cosmic rays 
is consistent with that of cosmic rays with lower energies 
and is uniform. 
It is noteworthy that 
three of seven 10$^{20}$eV cosmic rays are members of doublets. 
The BC1 cluster is in the direction of the Cygnus Loop or 
PSR 2053$+$36 region. 
It is desirable to examine any signals 
from this direction in other energy band around MJD 50,000. 
We hope other experiments in TeV -- PeV regions 
to explore the C1 -- C5 clusters and 10$^{20}$eV cosmic ray directions. 

%*********************************************************************
\acknowledgments

We are grateful to Akeno-mura, Nirasaki-shi, Sudama-cho, Nagasaka-cho, 
Ohizumi-mura, Tokyo Electric Power Co. and 
Nihon Telegram and Telephone Co. for their kind cooperation. 
The authors are indebted to other members of the Akeno group 
in the maintenance of the AGASA array. 
The authors are grateful to Prof. V. Berezinsky for 
his suggestions on the analysis of the Dark Matter Halo hypothesis. 
M.Takeda acknowledges the receipt of JSPS Research Fellowships. 
The authors thank Paul Sommers for his 
valuable suggestion on the preparation of the manuscript. 

%*********************************************************************

%*********************************************************************
%		Figure Captions
\clearpage

\figcaption[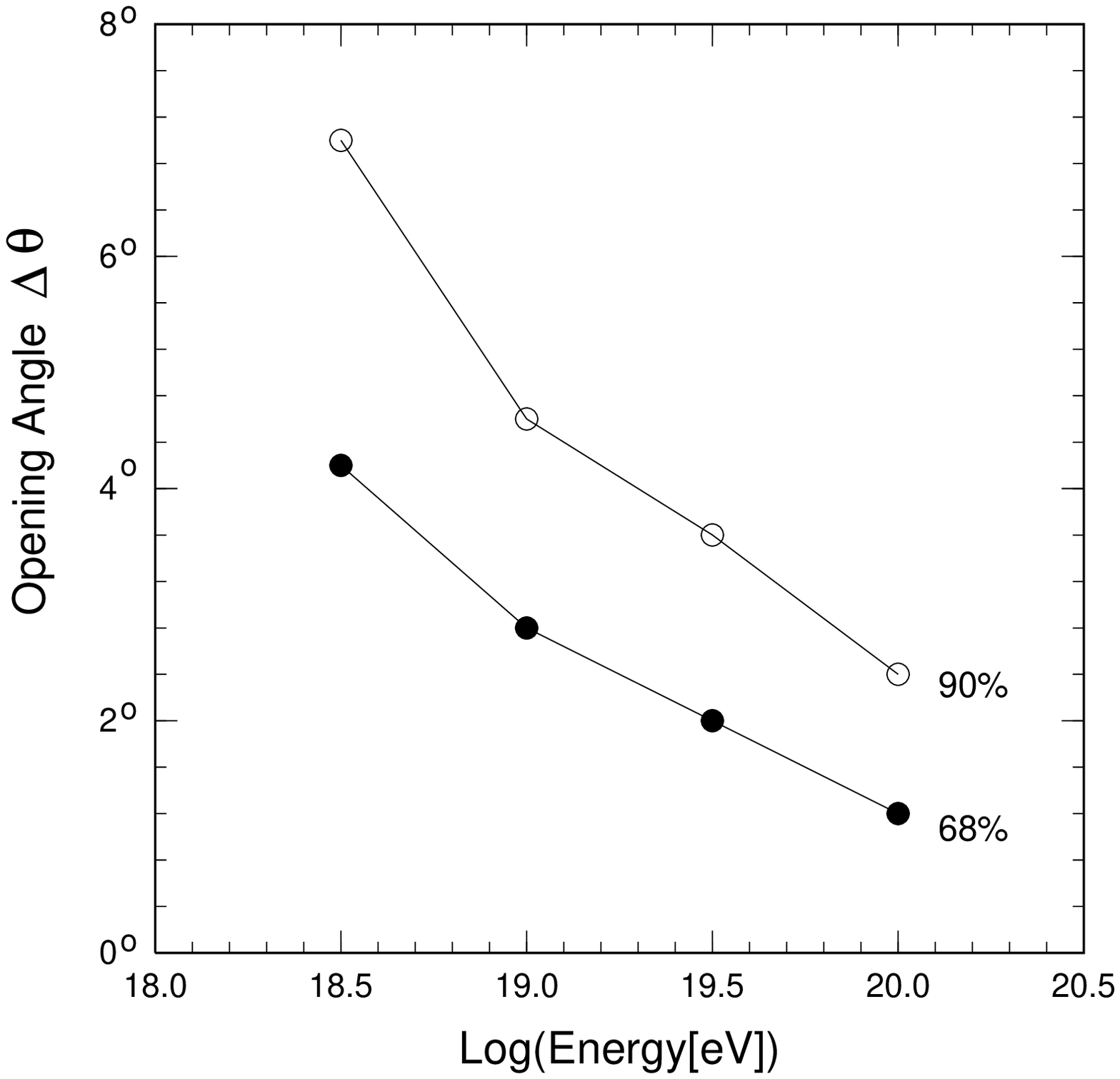]{
Accuracy on arrival direction determination. 
Closed and open circles are the opening angles 
encompassed 68 \% and 90 \% data. 
\label{fig:acc_dir}
}

\figcaption[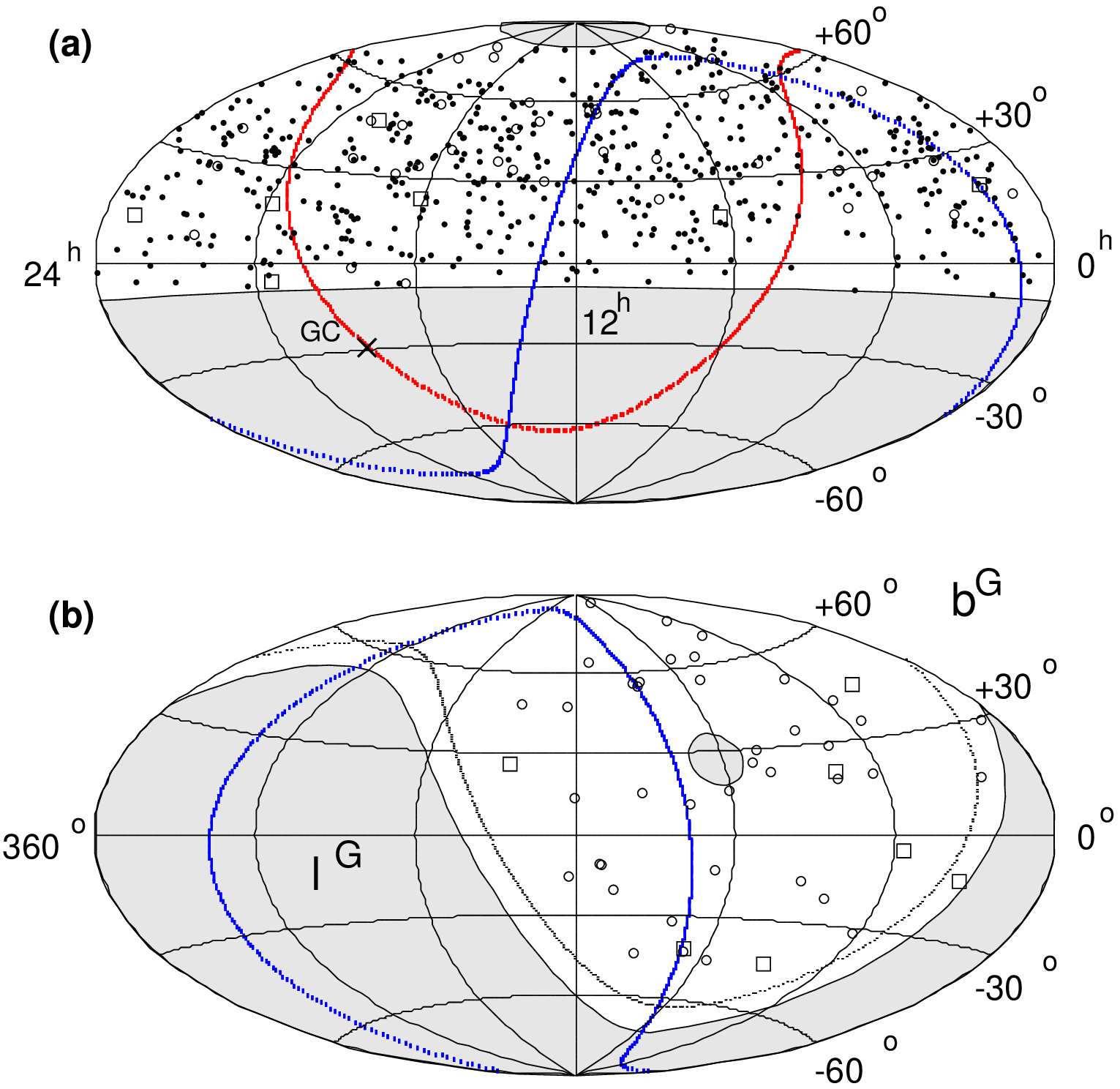]{
Arrival directions of cosmic rays with energies above 10$^{19.0}$eV 
on the (a) equatorial and (b) galactic coordinates. 
Dots, open circles, and open squares represent cosmic rays with energies of 
(1 -- 4) $\times$ 10$^{19}$eV, 
(4 -- 10) $\times$ 10$^{19}$eV, and $\geq$ 10$^{20}$eV, 
respectively. 
The galactic and supergalactic planes are shown by the dotted curves. 
``GC'' designates the galactic center.
\label{fig:equ1900}
}

\figcaption[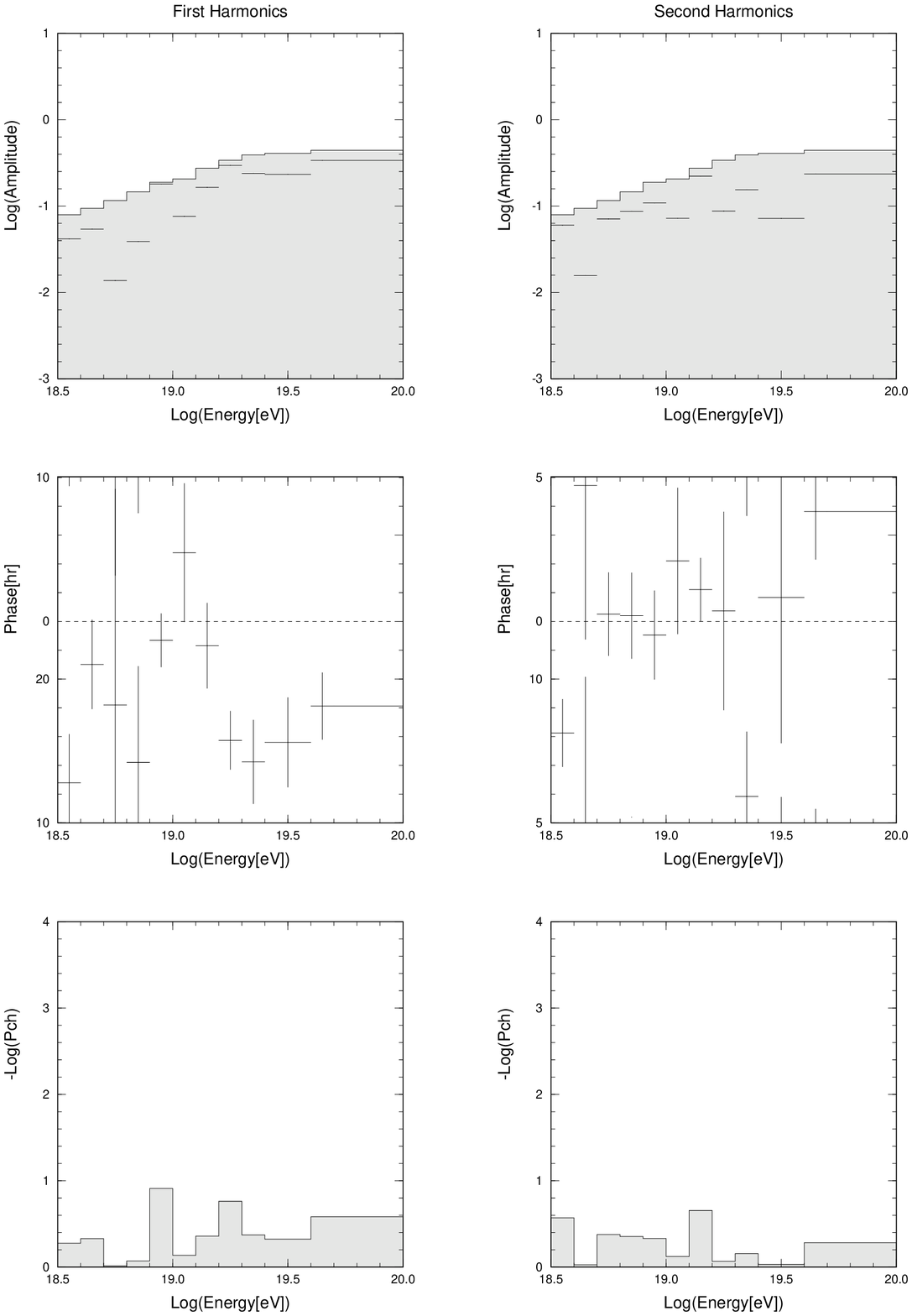]{
Results of the harmonic analysis. 
(Top to bottom, the amplitude, the phases and the chance probabilities 
of the first (left) and second (right) harmonics.)
\label{fig:harmo}
}

\figcaption[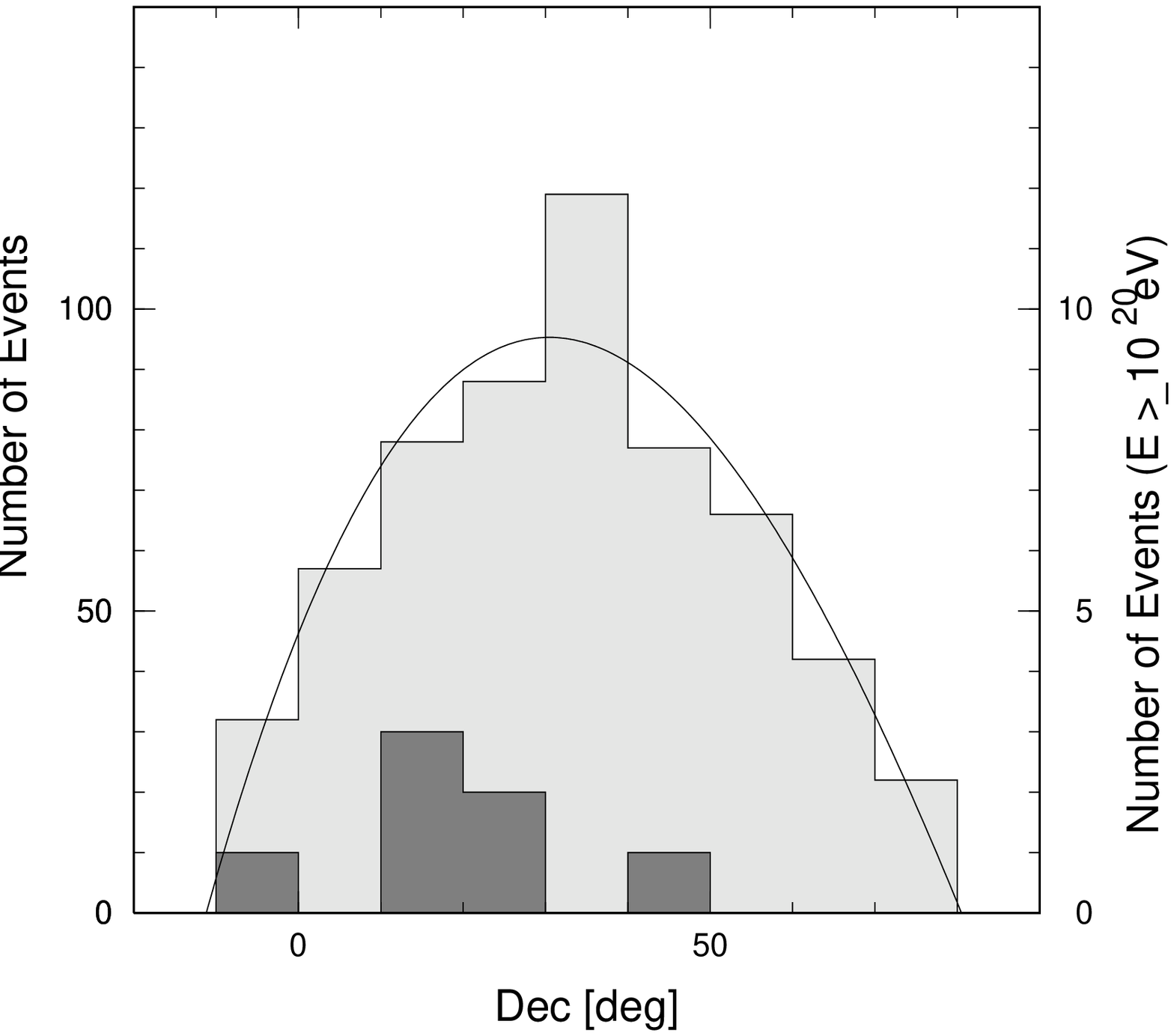]{
Declination distribution of the observed cosmic rays. 
(Light shaded histogram: $\geq$ 10$^{19}$eV. 
Dark shaded histogram: $\geq$ 10$^{20}$eV, 
the right-had vertical axis should be referred.)
\label{fig:dec1900}
}

\figcaption[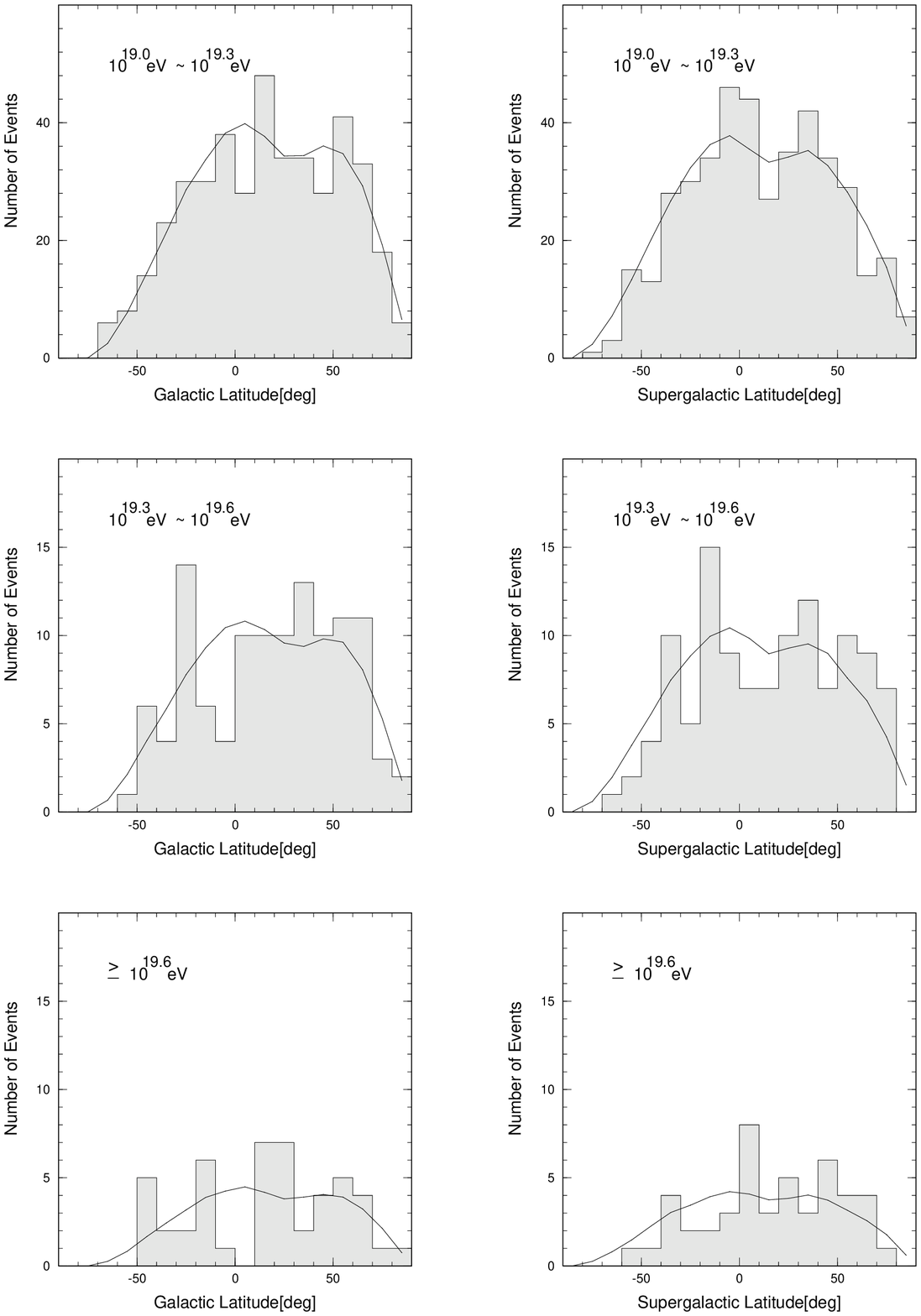]{
Galactic (left) and supergalactic (right) latitude distribution.
(Top:   (1 -- 2) $\times$ 10$^{19}$eV. 
Middle: (2 -- 4) $\times$ 10$^{19}$eV.
Bottom: $\geq$ 4 $\times$ 10$^{19}$eV.)
\label{fig:lat_gsg}
}

\figcaption[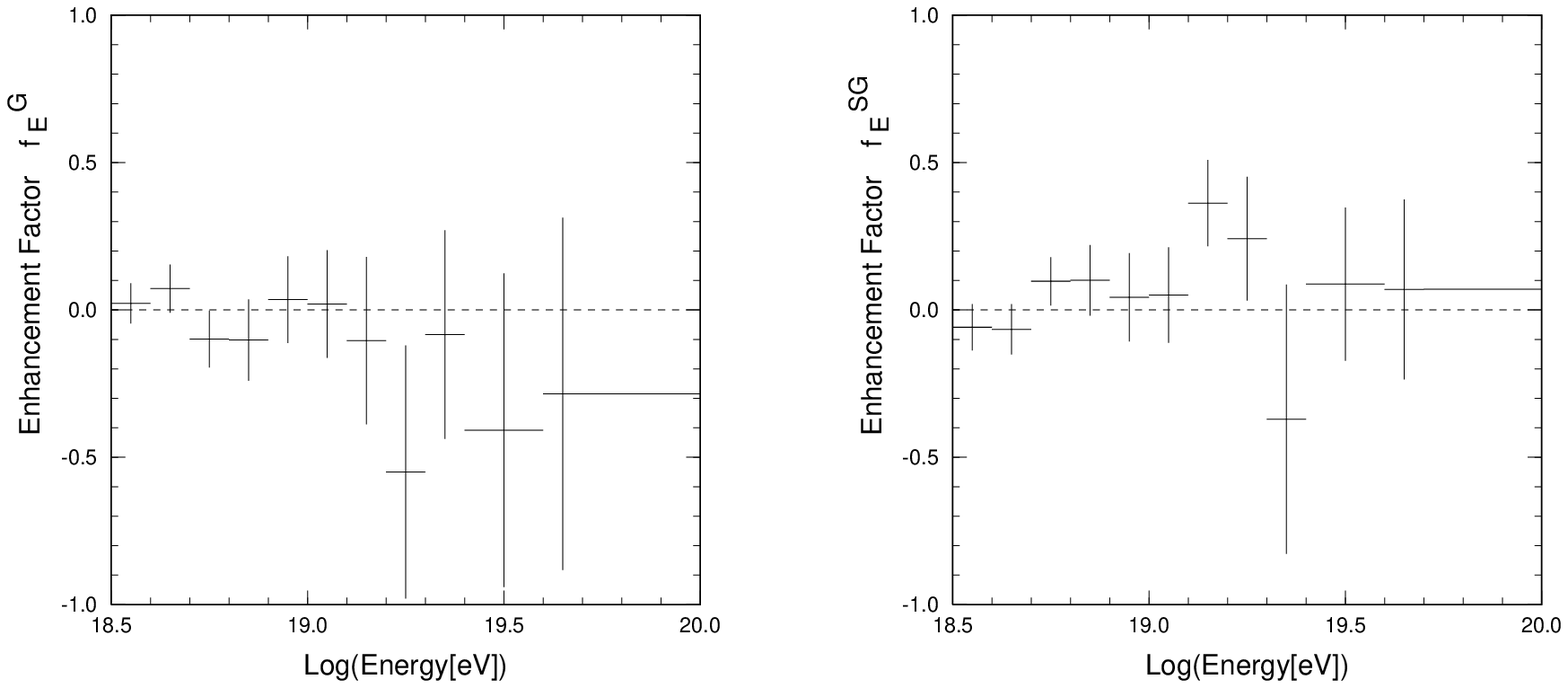]{
Dependence of the plane enhancement factor on the energy.
(Left: for the galactic coordinates. 
Right: for the supergalactic coordinates)
\label{fig:fe_gsg}
}

\figcaption[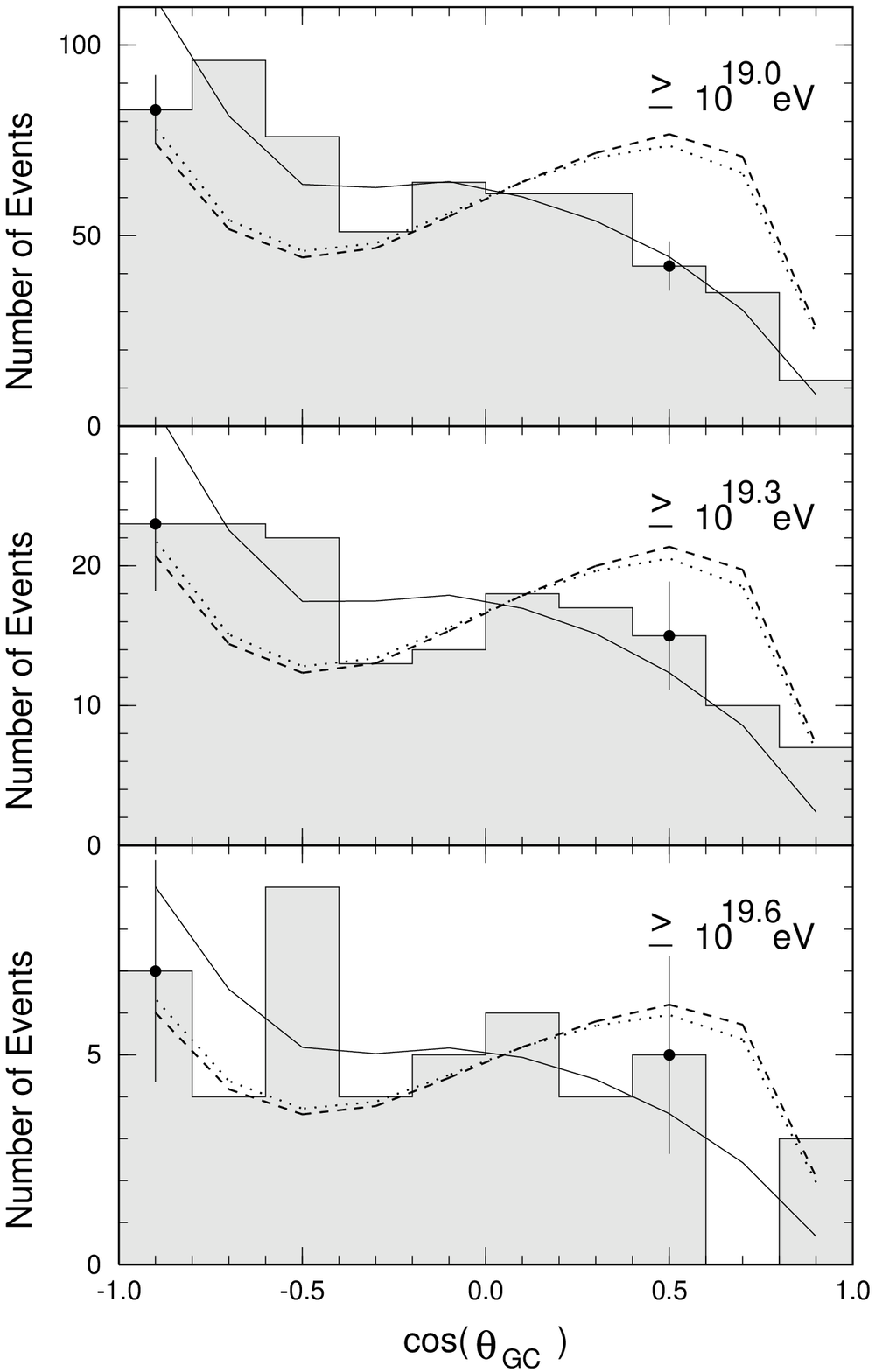]{
$\cos(\theta_{GC})$ distribution. 
(Top: $\geq$ 10$^{19}$eV. Middle: $\geq$ 2 $\times$ 10$^{19}$eV. 
Bottom: $\geq$ 4 $\times$ 10$^{19}$eV.)
Here, $\theta_{GC}$ is the opening angle between 
the cosmic-ray direction and the galactic center direction, 
with energies $\geq$ 10$^{19}$eV (top), 
2 $\times$ 10$^{19}$eV (middle), and 4 $\times$ 10$^{19}$eV (bottom). 
The solid, dashed and dotted curves indicate 
the distribution expected for the isotropic, ISO and NFW 
models, respectively. 
\label{fig:axhist_GC}
}

\figcaption[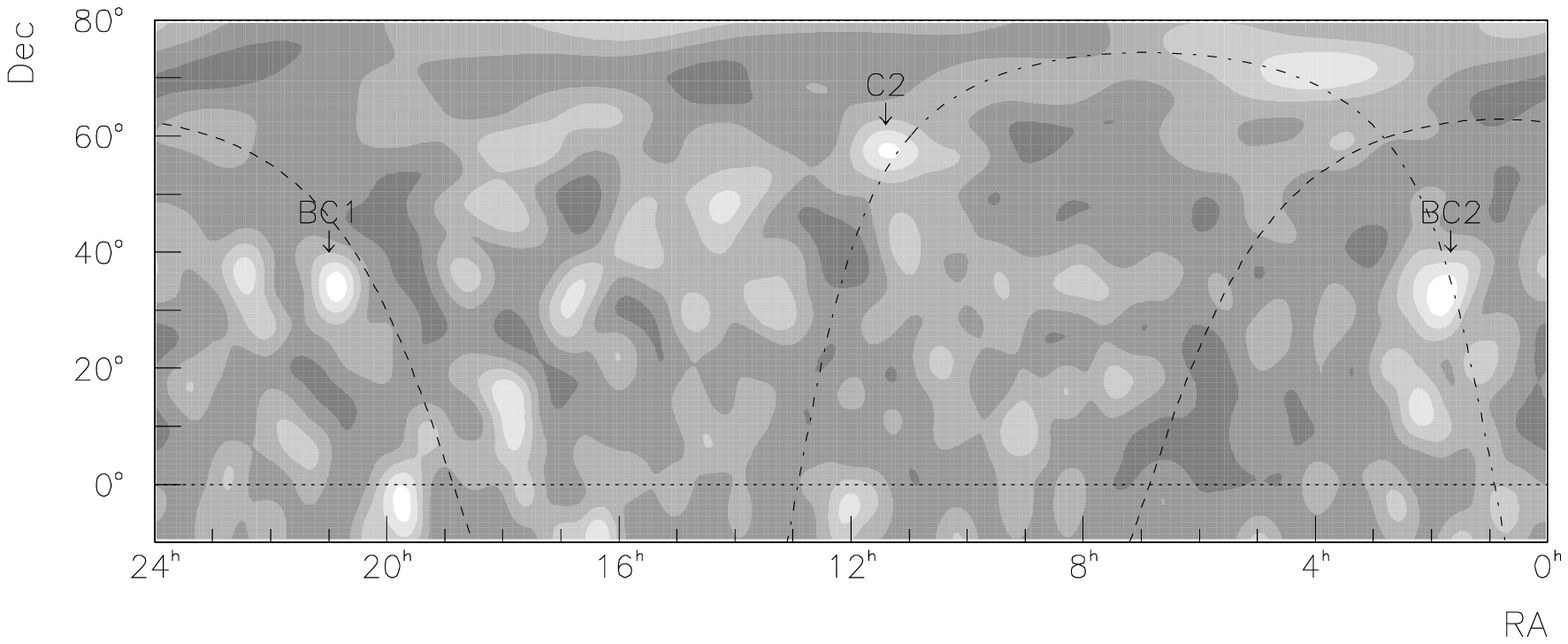]{
Significance map of cosmic-ray excess/deficit above 10$^{19}$eV.
The dashed and dash-dotted curve indicate 
the galactic and supergalactic plane, respectively. 
\label{fig:dmap1900}
}

\figcaption[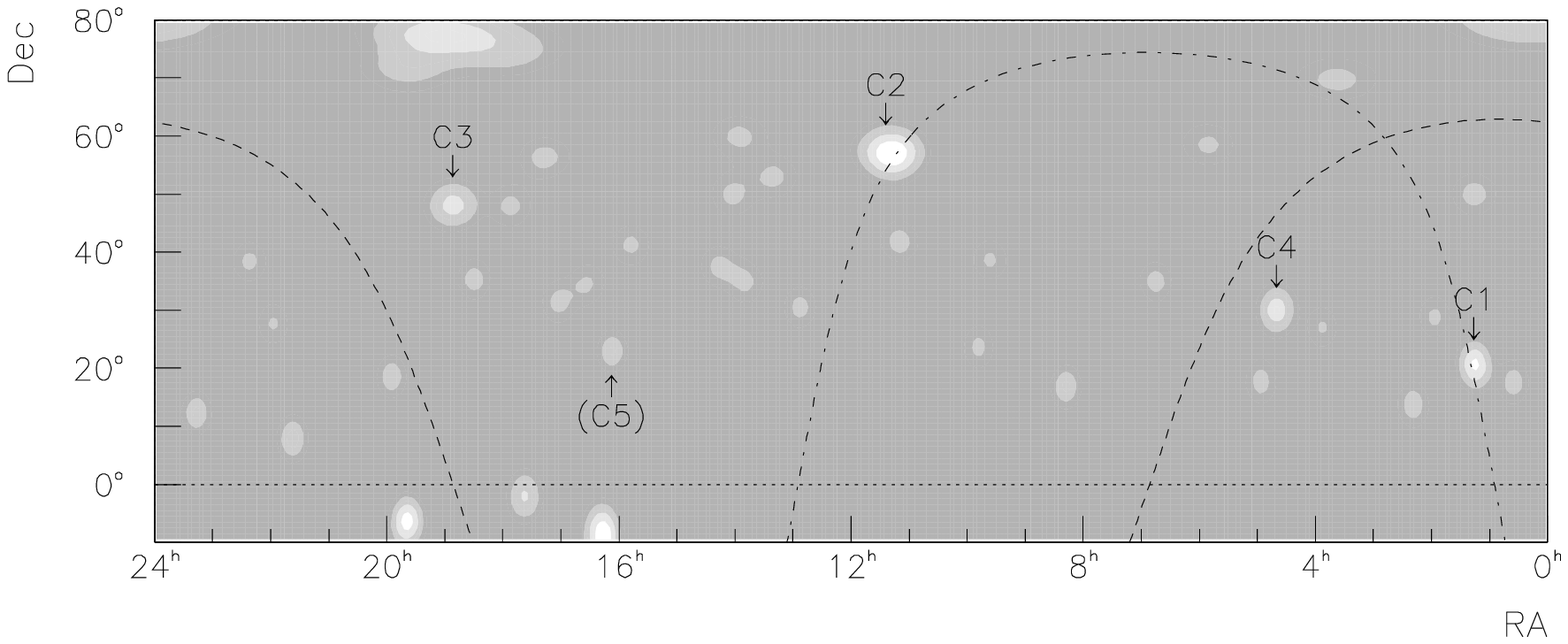]{
Significance map of cosmic-ray excess/deficit above 4 $\times$ 10$^{19}$eV.
The dashed and dash-dotted curve indicate 
the galactic and supergalactic plane, respectively. 
\label{fig:dmap1960}
}

\figcaption[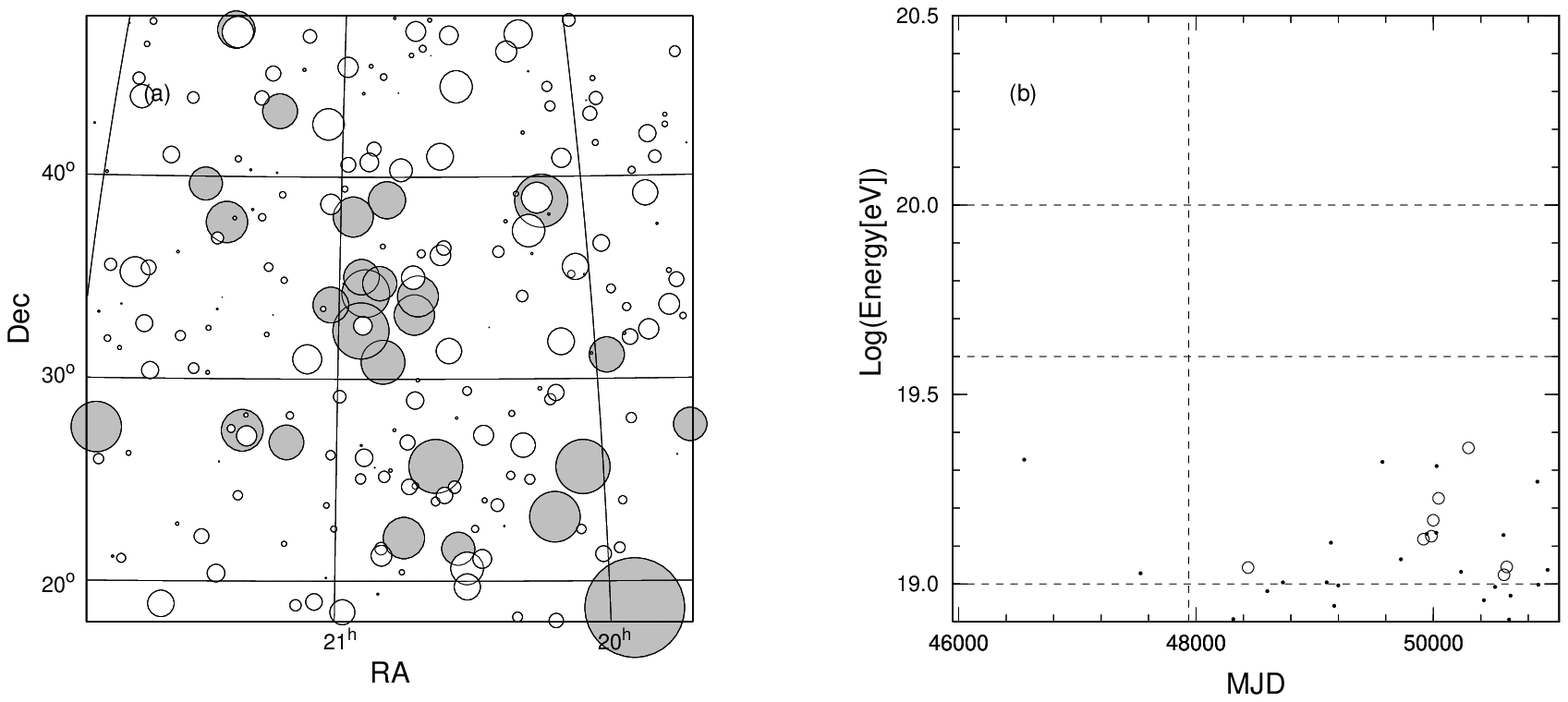]{
BC1 cluster. 
(a) Arrival directions of cosmic rays around the BC1 cluster. 
Radius of each circle corresponds to $\log$(E[eV]), and 
shaded and open circles have energies above 10$^{19}$eV and 
between 3 $\times$ 10$^{18}$eV and 10$^{19}$eV, respectively.
(b) Arrival time -- energy relation. 
Open circles denote members of the BC1 cluster and 
dots are cosmic rays near the BC1 cluster. 
After the vertical dotted line, A20 is combined into AGASA. 
\label{fig:clst_c0}
}

\figcaption[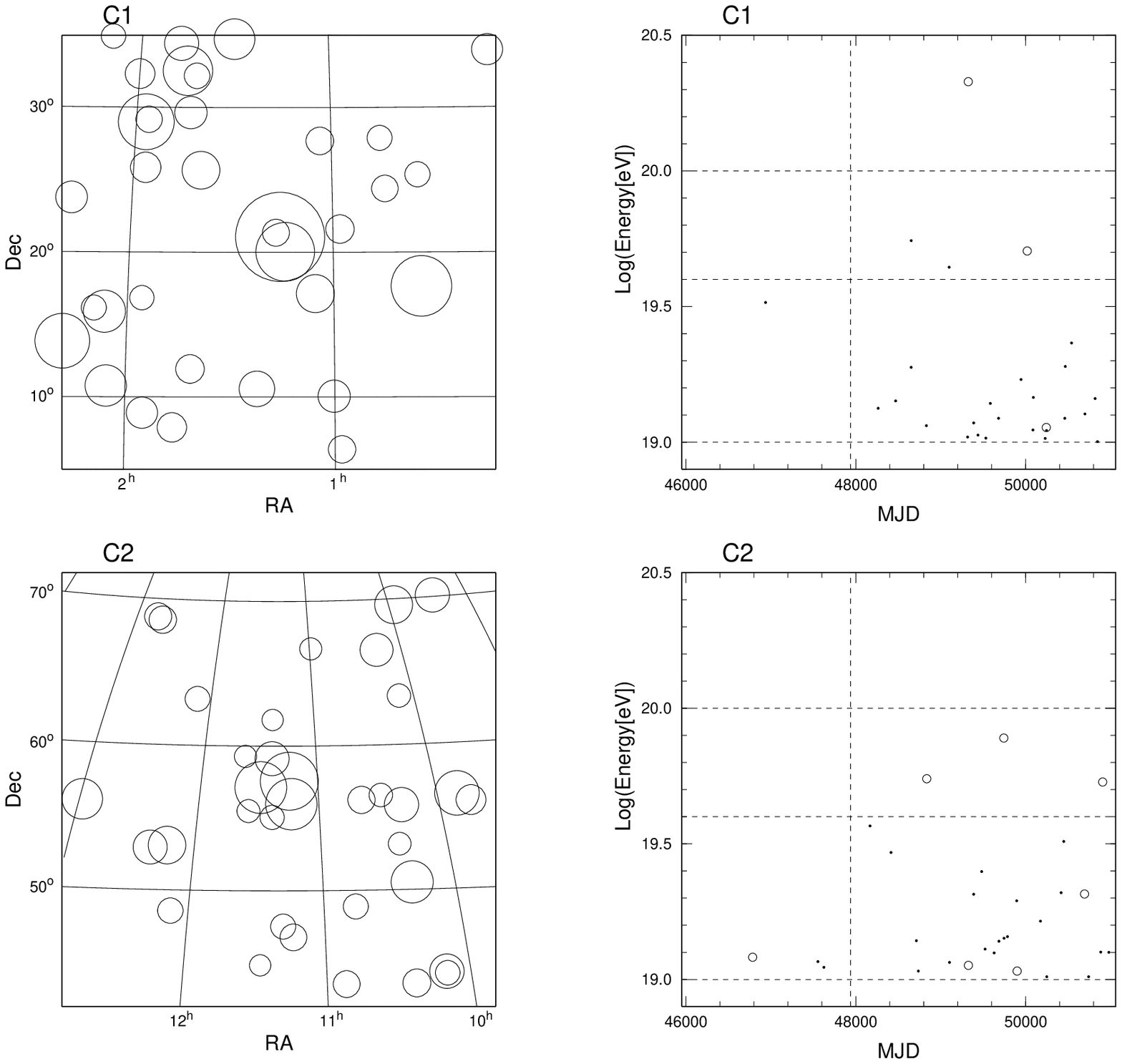]{
Arrival directions and arrival time -- energy relation 
for the C1 and C2 clusters. 
Here, cosmic rays above 10$^{19}$eV are plotted. 
(See also Figure 10)
\label{fig:clst_1960}
}

\figcaption[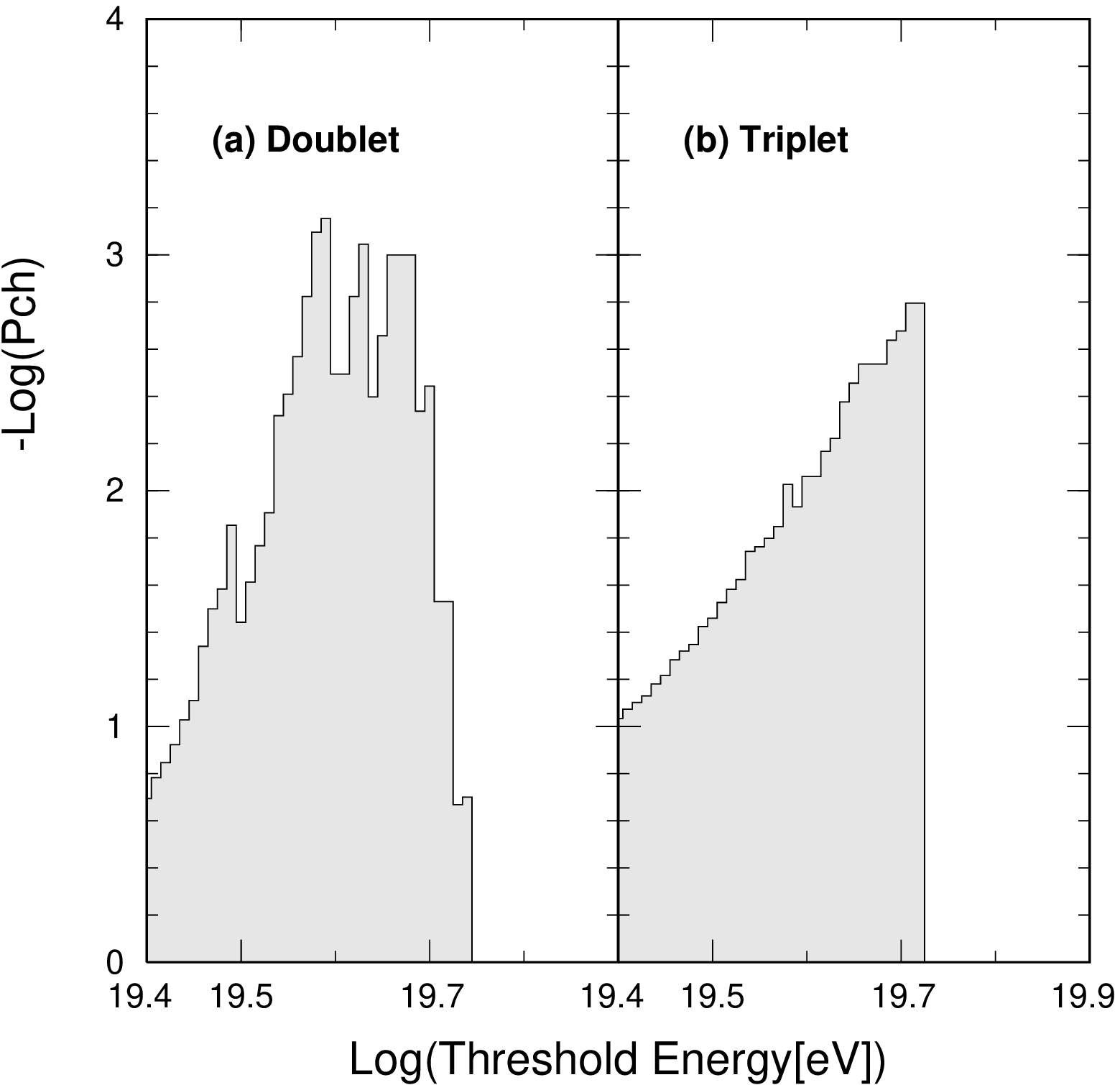]{
Energy dependence of the chance probability of observing 
(a) doublets and (b) triplets. 
\label{fig:clst_edep}
}

%*********************************************************************
%		Tables
\clearpage

\begin{deluxetable}{lrrr}
\tablecolumns{4}
%\footnotesize
%\tablewidth{0pc}
\tablenum{1}
\tablecaption{Number of events in the data set. \label{tbl:num_events}}
\tablehead{
\colhead{Array}				&
\colhead{$\geq$            10$^{19}$eV}	&
\colhead{$\geq$ 4 $\times$ 10$^{19}$eV}	&
\colhead{$\geq$            10$^{20}$eV}
}
\startdata
A20	&  59	&  7	& 0	\nl
AGASA	& 522	& 40	& 7	\nl
\tableline
Total	& 581	& 47	& 7	\nl
\enddata
\end{deluxetable}

\begin{deluxetable}{lllrrrrc}
\tablecolumns{8}
\footnotesize
%\tablewidth{0pc}
\tablenum{2}
\tablecaption{AGASA events 4 $\times$ 10$^{19}$eV. 
\label{tbl:40EeV}}
\tablehead{
\colhead{Date}			&
\colhead{Time(JST)}		&
\colhead{Energy}		&
\multicolumn{4}{c}{Coordinates\tablenotemark{1}}	&
\colhead{Note}			\\
\colhead{}			&
\colhead{}			&
\colhead{}			&
\colhead{$\alpha$}		&
\colhead{$\delta$}		&
\colhead{$l^{G}$}		&
\colhead{$b^{G}$}		&
\colhead{}
}
\startdata
 84/12/12 & 14:18:02 & 6.81 $\times$ 10$^{19}$ eV & 22$^{h}$ 21$^{m}$ & 38.4$^{\circ}$     &  93.3$^{\circ}$     & $-$15.7$^{\circ}$ &    \\
 84/12/17 & 10:28:16 & 9.79                        & 18$^{h}$ 29$^{m}$ & 35.3$^{\circ}$     &  63.5$^{\circ}$     &  19.4$^{\circ}$ &    \\
 86/01/05 & 19:31:03 & 5.47                        &  4$^{h}$ 38$^{m}$ & 30.1$^{\circ}$     & 170.4$^{\circ}$     & $-$11.2$^{\circ}$ & C4 \\
 86/10/23 & 14:25:15 & 6.22                        & 14$^{h}$ 02$^{m}$ & 49.9$^{\circ}$     &  96.8$^{\circ}$     &  63.4$^{\circ}$ &    \\
 87/11/26 & 17:49:20 & 4.82                        & 21$^{h}$ 57$^{m}$ & 27.6$^{\circ}$     &  82.1$^{\circ}$     & $-$21.1$^{\circ}$ &    \\
%     &      &        &          &          &         &         &    \\
 89/03/14 & 02:45:39 & 5.27                        & 13$^{h}$ 48$^{m}$ & 34.7$^{\circ}$     &  68.3$^{\circ}$     &  75.6$^{\circ}$ &    \\
 89/08/16 & 08:32:01 & 4.07                        &  5$^{h}$ 51$^{m}$ & 58.5$^{\circ}$     & 154.5$^{\circ}$     &  15.6$^{\circ}$ &    \\
 90/11/25 & 11:05:39 & 4.51                        & 16$^{h}$ 17$^{m}$ & $-$7.2$^{\circ}$     &   6.1$^{\circ}$     &  29.6$^{\circ}$ &    \\
 91/04/03 & 00:32:40 & 5.09                        & 15$^{h}$ 47$^{m}$ & 41.0$^{\circ}$     &  65.7$^{\circ}$     &  51.5$^{\circ}$ &    \\
 91/04/20 & 08:24:49 & 4.35                        & 18$^{h}$ 59$^{m}$ & 47.8$^{\circ}$     &  77.9$^{\circ}$     &  18.4$^{\circ}$ & C3 \\
     &      &        &          &          &         &         &    \\
 91/05/31 & 13:07:04 & 5.53                        &  3$^{h}$ 37$^{m}$ & 69.5$^{\circ}$     & 136.6$^{\circ}$     &  11.2$^{\circ}$ &    \\
 91/11/29 & 14:53:03 & 9.10                        & 19$^{h}$ 06$^{m}$ & 77.2$^{\circ}$     & 108.8$^{\circ}$     &  25.6$^{\circ}$ &    \\
 91/12/10 & 18:59:10 & 4.24                        &  0$^{h}$ 12$^{m}$ & 78.6$^{\circ}$     & 121.0$^{\circ}$     &  15.9$^{\circ}$ &    \\
 92/01/07 & 03:16:49 & 4.51                        &  9$^{h}$ 36$^{m}$ & 38.6$^{\circ}$     & 184.3$^{\circ}$     &  48.0$^{\circ}$ &    \\
 92/01/24 & 12:26:17 & 4.88                        & 17$^{h}$ 52$^{m}$ & 47.9$^{\circ}$     &  74.8$^{\circ}$     &  29.4$^{\circ}$ &    \\
%     &      &        &          &          &         &         &    \\
 92/02/01 & 17:20:52 & 5.53                        &  0$^{h}$ 34$^{m}$ & 17.7$^{\circ}$     & 117.2$^{\circ}$     & $-$45.0$^{\circ}$ &    \\
 92/03/30 & 03:05:30 & 4.47                        & 17$^{h}$ 03$^{m}$ & 31.4$^{\circ}$     &  53.6$^{\circ}$     &  35.6$^{\circ}$ &    \\
 92/08/01 & 13:00:47 & 5.50                        & 11$^{h}$ 29$^{m}$ & 57.1$^{\circ}$     & 143.2$^{\circ}$     &  56.6$^{\circ}$ & C2 \\
 92/09/13 & 08:59:44 & 9.25                        &  6$^{h}$ 44$^{m}$ & 34.9$^{\circ}$     & 180.5$^{\circ}$     &  13.9$^{\circ}$ &    \\
 93/01/12 & 02:41:13 &  \underline{10.1}\tablenotemark{2}      &  8$^{h}$ 17$^{m}$ & 16.8$^{\circ}$     & 206.7$^{\circ}$     &  26.4$^{\circ}$ &    \\
     &      &        &          &          &         &         &    \\
 93/01/21 & 07:58:06 & 4.46        & 13$^{h}$ 55$^{m}$ & 59.8$^{\circ}$     & 108.8$^{\circ}$     &  55.5$^{\circ}$ &    \\
 93/04/22 & 09:39:56 & 4.42                        &  1$^{h}$ 56$^{m}$ & 29.0$^{\circ}$     & 139.8$^{\circ}$     & $-$31.7$^{\circ}$ &    \\
 93/06/12 & 06:14:27 & 6.49                        &  1$^{h}$ 16$^{m}$ & 50.0$^{\circ}$     & 127.0$^{\circ}$     & $-$12.7$^{\circ}$ &    \\
 93/12/03 & 21:32:47 &  \underline{21.3}                        &  1$^{h}$ 15$^{m}$ & 21.1$^{\circ}$     & 130.5$^{\circ}$     & $-$41.4$^{\circ}$ & C1 \\
 94/07/06 & 20:34:54 &  \underline{13.4}                        & 18$^{h}$ 45$^{m}$ & 48.3$^{\circ}$     &  77.6$^{\circ}$     &  20.9$^{\circ}$ & C3 \\
\enddata
\tablenotetext{1}{The celestial coordinates are based 
on the J2000.0 coordinates.}
\tablenotetext{2}{The energies are re-evaluated after 
the system response have been checked in October 1997.}
\end{deluxetable}

\begin{deluxetable}{lllrrrrc}
\tablecolumns{8}
\footnotesize
%\tablewidth{0pc}
\tablenum{2}
\tablecaption{AGASA events above 4 $\times$ 10$^{19}$eV (continued).}
\tablehead{
\colhead{Date}			&
\colhead{Time(JST)}		&
\colhead{Energy}		&
\multicolumn{4}{c}{Coordinates}	&
\colhead{Note}			\\
\colhead{}			&
\colhead{}			&
\colhead{}			&
\colhead{$\alpha$}		&
\colhead{$\delta$}		&
\colhead{$l^{G}$}		&
\colhead{$b^{G}$}		&
\colhead{}
}
\startdata
 94/07/28 & 08:23:37 & 4.08 $\times$ 10$^{19}$ eV  &  4$^{h}$ 56$^{m}$ & 18.0$^{\circ}$     & 182.8$^{\circ}$     & $-$15.5$^{\circ}$ &    \\
 95/01/26 & 03:27:16 & 7.76                        & 11$^{h}$ 14$^{m}$ & 57.6$^{\circ}$     & 145.5$^{\circ}$     &  55.1$^{\circ}$ & C2 \\
 95/03/29 & 06:12:27 & 4.27                        & 17$^{h}$ 37$^{m}$ & $-$1.6$^{\circ}$     &  22.8$^{\circ}$     &  15.7$^{\circ}$ &    \\
 95/04/04 & 23:15:09 & 5.79                        & 12$^{h}$ 52$^{m}$ & 30.6$^{\circ}$     & 117.5$^{\circ}$     &  86.5$^{\circ}$ &    \\
 95/10/29 & 00:32:16 & 5.07                        &  1$^{h}$ 14$^{m}$ & 20.0$^{\circ}$     & 130.2$^{\circ}$     & $-$42.5$^{\circ}$ & C1 \\
%     &      &        &          &          &         &         &    \\
 95/11/15 & 04:27:45 & 4.89                        &  4$^{h}$ 41$^{m}$ & 29.9$^{\circ}$     & 171.1$^{\circ}$     & $-$10.8$^{\circ}$ & C4 \\
 96/01/11 & 09:01:21 &  \underline{14.4}                        & 16$^{h}$ 06$^{m}$ & 23.0$^{\circ}$     &  38.9$^{\circ}$     &  45.8$^{\circ}$ & C5 \\
 96/01/19 & 21:46:12 & 4.80                        &  3$^{h}$ 52$^{m}$ & 27.1$^{\circ}$     & 165.4$^{\circ}$     & $-$20.4$^{\circ}$ &    \\
 96/05/13 & 00:07:48 & 4.78                        & 17$^{h}$ 56$^{m}$ & 74.1$^{\circ}$     & 105.1$^{\circ}$     &  29.8$^{\circ}$ &    \\
 96/10/06 & 13:36:43 & 5.68                        & 13$^{h}$ 18$^{m}$ & 52.9$^{\circ}$     & 113.8$^{\circ}$     &  63.7$^{\circ}$ &    \\
     &      &        &          &          &         &         &    \\
 96/10/22 & 15:24:10 & \underline{10.5}                        & 19$^{h}$ 54$^{m}$ & 18.7$^{\circ}$     &  56.8$^{\circ}$     &  $-$4.8$^{\circ}$ &    \\
 96/11/12 & 16:58:42 & 7.46                        & 21$^{h}$ 37$^{m}$ &  8.1$^{\circ}$     &  62.7$^{\circ}$     & $-$31.3$^{\circ}$ &    \\
 96/12/08 & 12:08:39 & 4.30                        & 16$^{h}$ 31$^{m}$ & 34.6$^{\circ}$     &  56.2$^{\circ}$     &  42.8$^{\circ}$ &    \\
 96/12/24 & 07:36:36 & 4.97                        & 14$^{h}$ 17$^{m}$ & 37.7$^{\circ}$     &  68.5$^{\circ}$     &  69.1$^{\circ}$ &    \\
 97/03/03 & 07:17:44 & 4.39                        & 19$^{h}$ 37$^{m}$ & 71.1$^{\circ}$     & 103.0$^{\circ}$     &  21.9$^{\circ}$ &    \\
%     &      &        &          &          &         &         &    \\
 97/03/30 & 07:58:21 &  \underline{15.0}                        & 19$^{h}$ 38$^{m}$ & $-$5.8$^{\circ}$     &  33.1$^{\circ}$     & $-$13.1$^{\circ}$ &    \\
 97/04/28 & 13:46:18 & 4.20                        &  2$^{h}$ 18$^{m}$ & 13.8$^{\circ}$     & 152.9$^{\circ}$     & $-$43.9$^{\circ}$ &    \\
 97/11/20 & 07:23:25 & 7.21                        & 11$^{h}$ 09$^{m}$ & 41.8$^{\circ}$     & 171.2$^{\circ}$     &  64.6$^{\circ}$ &    \\
 98/02/06 & 00:12:26 & 4.11                        &  9$^{h}$ 47$^{m}$ & 23.7$^{\circ}$     & 207.2$^{\circ}$     &  48.6$^{\circ}$ &    \\
 98/03/30 & 08:17:26 & 6.93                        & 17$^{h}$ 16$^{m}$ & 56.3$^{\circ}$     &  84.5$^{\circ}$     &  35.3$^{\circ}$ &    \\
     &      &        &          &          &         &         &    \\
 98/04/04 & 20:07:03 & 5.35                        & 11$^{h}$ 13$^{m}$ & 56.0$^{\circ}$     & 147.5$^{\circ}$     &  56.2$^{\circ}$ & C2 \\
 98/06/12 & 06:43:49 &  \underline{12.0}                        & 23$^{h}$ 16$^{m}$ & 12.3$^{\circ}$     &  89.5$^{\circ}$     & $-$44.3$^{\circ}$ &     \\
 \hline
 97/04/10 & 02:48:48 & 3.89                        & 15$^{h}$ 58$^{m}$ & 23.7$^{\circ}$     &  39.1$^{\circ}$     &  47.8$^{\circ}$ & C5 \\
\enddata
\end{deluxetable}

\begin{deluxetable}{cllrrrr}
\tablecolumns{7}
%\footnotesize
%\tablewidth{0pc}
\tablenum{3}
\tablecaption{Members of the clustering events above 
about 10$^{19}$eV. 
\label{tbl:clst_1900}}
\tablehead{
\colhead{Name}			&
\colhead{Date}			&
\colhead{Energy}		&
\multicolumn{4}{c}{Coordinates\tablenotemark{1}}	\\
  &  &	& $\alpha$	& $\delta$	& $l^{G}$	& $b^{G}$
}
\startdata
BC1	& 95/10/09	& 1.47 $\times$ 10$^{19}$eV
			& 20$^{h}$ 50$^{m}$	&    30.8$\arcdeg$
			&  73.9$\arcdeg$	& $-$ 8.2$\arcdeg$	\nl
	& 95/11/23	& 1.68
			& 20$^{h}$ 54$^{m}$	&    34.2$\arcdeg$
			&  77.1$\arcdeg$	& $-$ 6.8$\arcdeg$	\nl
	& 95/07/18	& 1.31
			& 20$^{h}$ 42$^{m}$	&    33.2$\arcdeg$
			&  74.8$\arcdeg$	& $-$ 5.5$\arcdeg$	\nl
	& 95/09/24	& 1.33
			& 20$^{h}$ 41$^{m}$	&    34.1$\arcdeg$
			&  75.4$\arcdeg$	& $-$ 4.8$\arcdeg$	\nl
	& 91/07/02	& 1.10
			& 20$^{h}$ 55$^{m}$	&    35.1$\arcdeg$
			&  77.9$\arcdeg$	& $-$ 6.4$\arcdeg$	\nl
	& 96/08/02	& 2.29
			& 20$^{h}$ 55$^{m}$	&    32.4$\arcdeg$
			&  75.9$\arcdeg$	& $-$ 8.1$\arcdeg$	\nl
	& 97/05/28	& 1.06
			& 20$^{h}$ 50$^{m}$	&    34.7$\arcdeg$
			&  77.1$\arcdeg$	& $-$ 5.9$\arcdeg$	\nl
	& 97/06/20	& 1.11
			& 21$^{h}$ 02$^{m}$	&    33.7$\arcdeg$
			&  77.8$\arcdeg$	& $-$ 8.4$\arcdeg$	\nl
\tableline
BC2	& 		& 
			&  1$^{h}$ 40$^{m}$	&    35$\arcdeg$
			& 134$\arcdeg$		& $-$27$\arcdeg$	\nl
\nl
\enddata
\tablenotetext{1}{The celestial coordinates are based 
on the J2000.0 coordinates.}
\end{deluxetable}

\begin{deluxetable}{lcclc}
\tablecolumns{5}
\footnotesize
%\tablewidth{0pc}
\tablenum{4}
\tablecaption{Kolmogorov-Smirnov test for celestial coordinates.
\label{tbl:100EeV_KS}}
\tablehead{
\colhead{}			&
\colhead{KS-Probability}	&
\colhead{\hspace{0.2em}}	&
\colhead{}			&
\colhead{KS-Probability}
}
\startdata
Azimuth Angle ($\phi$)			& 0.268		& &	% 0.26801
Zenith Angle ($\theta$)			& 0.867		\nl	% 0.86666
Right Ascension ($\alpha$)		& 0.202		& &	% 0.20241
Declination ($\delta$)			& 0.025		\nl	% 0.02526
Ecliptic Longitude			& 0.085		& &	% 0.08580
Ecliptic Latitude			& 0.449		\nl	% 0.44903
Galactic Longitude ($l^{G}$)		& 0.182		& &	% 0.18224
Galactic Latitute ($b^{G}$)		& 0.540		\nl	% 0.54014
Supergalactic Longitude ($l^{SG}$)	& 0.654		& &	% 0.65351
Supergalactic Latitude ($b^{SG}$)	& 0.167		\nl	% 0.16726
\enddata
\end{deluxetable}

\begin{deluxetable}{crrr}
\tablecolumns{4}
\footnotesize
%\tablewidth{0pc}
\tablenum{5}
\tablecaption{Reduced-$\chi^{2}$ values of the $\cos(\theta_{GC})$ 
distribution with three models. \label{tbl:gal_halo}}
\tablehead{
\colhead{}				&
\colhead{$\geq$ 10$^{19}$eV}		&
\colhead{$\geq$ 2 $\times$ 10$^{19}$eV}	&
\colhead{$\geq$ 4 $\times$ 10$^{19}$eV}	
}
\startdata
isotropic distribution	& 2.0	& 1.7	& 1.8	\nl
ISO model		& 11.8	& 2.2	& 1.7	\nl
NFW model		& 10.0	& 1.9	& 1.6	\nl
\enddata
\end{deluxetable}

\begin{deluxetable}{ll}
\tablecolumns{2}
\footnotesize
%\tablewidth{0pc}
\tablenum{6}
\tablecaption{Astrophysical objects near the AGASA events. 
\label{tbl:correlate}}
\tablehead{
\colhead{Event ID}		&
\colhead{Astrophysical object}
}
\startdata
%BC2	& Cygnus Loop, \, PSR 2053$+$36		\nl
C1	& Mrk 359 (z $=$ 0.017)			\nl
C2	& NGC 3642 (z $=$ 0.005), \, Mrk 40 (z $=$ 0.02),
	 \, Mrk 171 (z $=$ 0.01)		\nl
970330 (1.5 $\times$ 10$^{20}$eV)
	& H 1934$-$063 (z $=$ 0.011)		\nl
\enddata
\end{deluxetable}

%*********************************************************************
\clearpage
%		Figures

\begin{figure}
\figurenum{1}
\epsfxsize=\textwidth
\epsfbox{fig01.eps}
\caption{}
\end{figure}

\begin{figure}
\figurenum{2}
\epsfxsize=\textwidth
\epsfbox{fig02.eps}
\caption{}
\end{figure}

\begin{figure}
\figurenum{3}
\epsfysize=\textheight
\epsfbox{fig03.eps}
\caption{}
\end{figure}

\begin{figure}
\figurenum{4}
\epsfxsize=\textwidth
\epsfbox{fig04.eps}
\caption{}
\end{figure}

\begin{figure}
\figurenum{5}
\epsfysize=\textheight
\epsfbox{fig05.eps}
\caption{}
\end{figure}

\clearpage

\begin{figure}
\figurenum{6}
\epsfxsize=\textwidth
\epsfbox{fig06.eps}
\caption{}
\end{figure}

\clearpage

\begin{figure}
\figurenum{7}
\epsfysize=\textheight
\epsfbox{fig07.eps}
\caption{}
\end{figure}

\clearpage

\begin{figure}
\figurenum{8}
\epsfxsize=\textwidth
\epsfbox{fig08n.eps}
\caption{}
\end{figure}

\begin{figure}
\figurenum{9}
\epsfxsize=\textwidth
\epsfbox{fig09n.eps}
\caption{}
\end{figure}

\clearpage

\begin{figure}
\figurenum{10}
\epsfxsize=\textwidth
\epsfbox{fig10.eps}
\caption{}
\end{figure}

\begin{figure}
\figurenum{11}
\epsfxsize=\textwidth
\epsfbox{fig11.eps}
\caption{}
\end{figure}

\clearpage

\begin{figure}
\figurenum{12}
\epsfxsize=\textwidth
\epsfbox{fig12.eps}
\caption{}
\end{figure}

%*********************************************************************
\end{document}